\documentclass[twocolumn,showpacs,preprintnumbers,amsmath,amssymb,superscriptaddress]{revtex4-2}

\usepackage{epsf}
\usepackage{graphicx}
\usepackage{sidecap}
 \usepackage{soul}
 \usepackage{array}
 \usepackage{amsmath}
 \usepackage{amssymb}
 \usepackage{dcolumn}
 \usepackage{epstopdf}
 \usepackage{bm}
 \usepackage{amsmath}

\usepackage{color}
\usepackage{hyperref}
\usepackage{soul}
\sethlcolor{green}
\usepackage{bbding}
\hypersetup{
    colorlinks=true,
    citecolor=red,
    linkcolor=red,
    filecolor=blue,   
    urlcolor=blue,
}

\def \beq {\begin{equation}}
\def \eeq {\end{equation}}
\begin{document}

\textcolor{white}{!TeX document-id = {9c0ae04a-bcf3-4a29-a29d-35e056ca3658}
!TEX TS-program = pdflatexmk}

\title{Complex Fermiology and Electronic Structure of Antiferromagnet EuSnP}

\author{Milo Sprague} \thanks{These authors contributed equally in this work.} \affiliation {Department of Physics, University of Central Florida, Orlando, Florida 32816, USA}
\author{Anup Pradhan Sakhya} \thanks{These authors contributed equally in this work.} \affiliation {Department of Physics, University of Central Florida, Orlando, Florida 32816, USA}  
\author{Sabin~Regmi} \affiliation{Department of Physics, University of Central Florida, Orlando, Florida 32816, USA}
\author{Mazharul Islam Mondal} \affiliation{Department of Physics, University of Central Florida, Orlando, Florida 32816, USA} 
\author{Iftakhar Bin Elius} \affiliation{Department of Physics, University of Central Florida, Orlando, Florida 32816, USA}
\author{Nathan Valadez} \affiliation {Department of Physics, University of Central Florida, Orlando, Florida 32816, USA}
\author{Kali Booth} \affiliation {Department of Physics, University of Central Florida, Orlando, Florida 32816, USA}
\author{Tetiana Romanova} \affiliation{Institute of Low Temperature and Structure Research, Polish Academy of Sciences, ul. Okólna 2, 50-422 Wrocław, Poland}
\author{Andrzej~Ptok}\affiliation{Institute of Nuclear Physics, Polish Academy of Sciences, W. E. Radzikowskiego 152, PL-31342 Krak\'{o}w, Poland}
\author{Dariusz Kaczorowski}\affiliation{Institute of Low Temperature and Structure Research, Polish Academy of Sciences, ul. Okólna 2, 50-422 Wrocław, Poland}
\author{Madhab~Neupane} \thanks{corresponding author: Madhab.Neupane@ucf.edu} \affiliation {Department of Physics, University of Central Florida, Orlando, Florida 32816, USA}
\date{\today}

\begin{abstract}
\noindent We studied the electronic structure of a layered antiferromagnetic metal, EuSnP, in the paramagnetic and in the antiferromagnetic phase using angle-resolved photoemission spectroscopy (ARPES) alongside density functional theory (DFT) based first-principles calculations. The temperature dependence of the magnetic susceptibility measurements exhibits an antiferromagnetic transition at a N\'eel temperature of 21 K. Employing high-resolution ARPES, the valence band structure was measured at several temperatures above and below the N\'eel temperature, which produced identical spectra independent of temperature. Through analysis of the ARPES results presented here, we attribute the temperature-independent spectra to the weak coupling of the Sn, and P conduction electrons with Eu 4\textit{f} states. 
{\noindent }
\end{abstract}

\maketitle 
\indent The involvement of antiferromagnetic (AFM) orderings on the electronic band structure of materials have seen increased interest recently \cite{sobota, Hao2013FePn, Noh2014AFM, Kuroda, Sakhyasmbi, kaminskinature, Sakhya, kushnirenkindsb, SabinEuIn2As2}. Antiferromagnetism, where the net magnetization vanishes due to compensation of magnetic moments, is capable of producing a variety of ordering forms that modify the crystal symmetry experienced by itinerant electrons \cite{Dress}. This can include a redefining of the unit-cell periodicity, which results in the reduction of the Brillouin zone (BZ) in reciprocal space. Such re-scaling of the BZ induces band-folding, from which band-hybridization may follow, dramatically reconstructing the electronic structure \cite{sobota, Hao2013FePn, Noh2014AFM, Kuroda, Sakhya}. Another way in which AFM ordering can modify the system is through changing the crystal space-group \cite{Cracknell, Burns}. Rotational and mirror symmetries may be broken by the alignment of magnetic moments, which may remove degeneracies and forced band dispersions along high symmetry directions of the BZ \cite{Birss}.\newline 
\indent EuSnP is an AFM material which crystallizes in the NbCrN-type primitive tetragonal crystal structure (P4/nmm, No. 129) with six atoms in each unit cell and lattice constants of  a = b= 4.27 $\text{\AA}$  and c = 8.76 $\text{\AA}$ \cite{Payne, Iha}. It is a layered compound with Eu-Sn-P and P-Sn-Eu linear chains along the tetragonal [001] axis \cite{Iha}. Magnetic susceptibility measurements on single crystals of EuSnP as a function of temperature indicate a paramagnetic (PM) to AFM transition with a N\'{e}el temperature of 21 K \cite{Payne}. It is observed that the magnetic susceptiblity along the \textit{c}-axis ($\chi_{c}$) is suppressed below the N\'{e}el temperature when compared with the magnetic susceptibility along the \textit{a}-axis ($\chi_{a}$), exhibiting an anisotropic behavior \cite{Fujiwara}. The PM Curie temperatures estimated from $\chi_{c}^{-1}(T)$ and $\chi_{a}^{-1}(T)$ curves respectively are 13.3 K and 18.2 K which suggests that the exchange coupling between the Eu sublayer is ferromagnetic and these ferromagnetic sublayers couple antiferromagnetically along the \textit{c}-axis \cite{Fujiwara}. Pressure studies on EuSnP observed no structural phase transition up to $\sim$ 6.2 GPa \cite{Gui}. High-pressure resistivity measurements suggested that the N\'eel temperature is significantly enhanced under high pressure up to 2.15 GPa \cite{Gui}. Most relevant to this work, angular-dependent de-Haas van-Alphen (dHvA) measurements of the valence electronic structure for EuSnP within the AFM phase have been performed, which presents the low-temperatue bulk Fermi pockets \cite{Iha}. The sample conditions under dHvA measurements produce the bulk Fermi pockets of EuSnP within the magnetically ordered phase, extending beyond the metamagnetic transitions. However, due to the presence of both a PM-AFM phase transition at 21 K and metamagnetic transitions at 1.9 T and 6.7 T \cite{Fujiwara}, field-free temperature dependent study of the electronic structure is necessary.

\indent In this Letter, we have performed a detailed study of EuSnP using high-resolution angle-resolved photoemission spectroscopy (ARPES) in both the PM and the AFM phase to understand the electronic structure of this material and the influence of AFM ordering on the electronic structure below and above the N\'eel 
 temperature. Our temperature dependent ARPES data suggest no significant electronic reconstruction across the AFM transition thus suggesting weak coupling of Eu 4\textit{f} electrons with Sn and P conduction electrons. Our density functional theory (DFT)-based first principles calculations show excellent agreement with the observed ARPES spectra. 
\begin{figure*}
	\includegraphics[width=16cm]{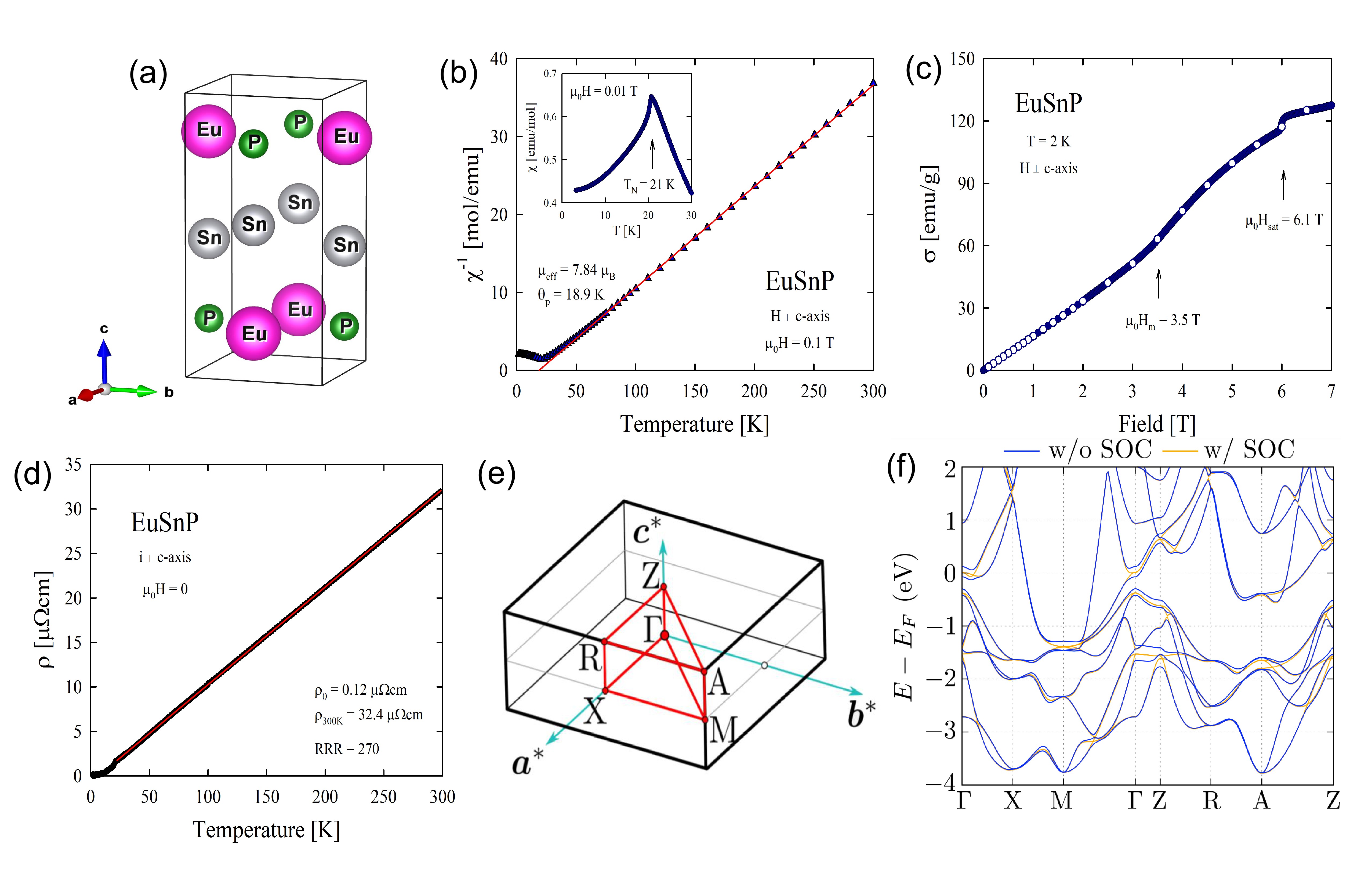}
\caption{Crystal structure and DFT calculations of EuSnP. (a) Tetragonal unit cell of EuSnP. Magenta, gray, and green colored solid spheres denote the Eu, Sn and P atoms, respectively. (b) Temperature dependence of the inverse magnetic susceptibility of EuSnP measured in a magnetic field of 0.1 T applied perpendicular to the tetragonal axis. Solid straight line represents the Curie-Weiss fit with the parameters given in the panel. Inset: low-temperature magnetic susceptibility data collected in a small field of 10 mT. Arrow marks the AFM transition at $T_N$. (c) Magnetic field variation of the magnetization in EuSnP taken at 2 K in magnetic fields directed perpendicular to the tetragonal axis. Arrows indicate a metamagnetic transition at $\mu_0 H_m$ and the onset of spin-polarized state above $H_{sat}$. (d) Temperature dependence of the electrical resistivity measured within the tetragonal plane. Solid straight line emphasizes a linear-in-T behavior in the PM phase. The measured crystal was characterized by a large value of the residual resistivity ratio (RRR) given in the panel. (e) Bulk Brillouin zone reciprocal to the tetragonal lattice of EuSnP. High symmetry points and reciprocal lattice vectors are indicated. (f) Electronic band structure calculations along various high-symmetry directions without (yellow curves) and with (blue curves) the inclusion of spin-orbit coupling.}
\label{fig1}
\end{figure*}

\indent High-quality single crystals of EuSnP were grown by Sn-flux method \cite{Iha}, and characterized by x-ray diffraction (Oxford Diffraction X'calibur four-circle diffractometer with a CCD Atlas detector) and energy dispersive x-ray analysis (FEI scanning electron
microscope with an EDAX Genesis XM4 spectrometer) to have the expected crystal structure and chemical composition. Their magnetic and electrical properties were verified by magnetic susceptibility (Quantum Design MPMS-XL magnetometer) and electrical resistivity (Quantum Design PPMS-14 platform) measurements. The ARPES measurements were performed at Stanford Synchrotron Radiation Lightsource (SSRL) endstation 5-2 and the ALS beamline 4.0.3. Measurements were carried out at temperatures of 18.5 K, 30 K, and 50 K. The pressure in the UHV chamber was maintained better than 1$\times$10$^{-10}$ torr. The angular and energy resolution were set better than 0.2$^\circ$ and 15 meV, respectively. Measurements were performed employing the photon energies in the range of 45 eV - 60 eV. DFT calculations were performed using the projector augmented-wave (PAW) potentials \cite{bloch} implemented in the Vienna Ab initio Simulation Package \cite{kresse, kresse1,kresse2}. Calculations are made within the generalized gradient approximation (GGA) in the Perdew, Burke, and Ernzerhof (PBE) parameterization \cite{perdew}. The energy cutoff for the plane-wave expansion was set to $400$~eV. In calculation we used the experimental lattice vectors as well as atomic positions \cite{Payne}. The electronic band structure was also evaluated within Quantum ESPRESSO \cite{giannozzi,giannozzi1,giannozzi2} with {\sc PsLibrary}~\cite{dal}. Exact results of the electronic band structure calculation, performed for primitive unit cell, was used to find the tight binding model in the basis of the maximally localized Wannier orbitals~\cite{marzari,marzari1,souza}. It was performed using the {\sc Wannier90} software~\cite{mostofi,mostofi1,pizzi}. In our calculations, we used the $8 \times 8 \times 8$ full ${\bm k}$-point mesh. Finally, $24$-orbital tight binding model of  EuSnP, was used to investigate the surface Green's function for semi-infinite system~\cite{sancho}, using {\sc WannierTools}~\cite{wu} software.

\indent The crystal structure of EuSnP is shown in Fig 1(a). The compound crystallizes in the tetragonal structure (P4/nmm, space group No. 129), which is in agreement with previous reports \cite{Iha, Payne}. The primitive unit cell contains Eu atoms at the $2c$ ($0$,$0.5$,$0.169$), Sn atoms at $4d$ ($0.5$,$0$,$0.45$), and P atoms at $4e$ ($0$,$0.5$,$0.84$)~\cite{Iha} Wyckoff positions. The crystal structure is composed of three layers which are stacked along the \textit{c}-axis where each Sn atom shares four equal bonds with other Sn atoms. The top and the bottom layer is connected via P and Eu atoms. Along the \textit{c}-direction, Eu atoms are arranged linearly with Eu-P-Sn-Eu order. Fig. 1(b) displays the inverse magnetic susceptibility of single-crystalline EuSnP measured in a magnetic field oriented perpendicular to the crystallographic \textit{c}-axis. Above 40 K, the $\chi^{-1}(T)$ dependence can be described by the Curie-Weiss law with the effective magnetic moment $\mu_{eff}$ = 7.84(2) $\mu_B$ and the PM Curie temperature of $\theta_p$ = 18.9(4) K. The value of $\mu_{eff}$ is close to the theoretical value
g[J(J+1)] 1/2 = 7.94 (Land\'e factor g = 2, total angular momentum J = 7/2) expected within Russell-Saunders coupling scheme for a divalent Eu ion. In turn, the positive value of $\theta_p$ hints at predominance of ferromagnetic exchange interactions. It is worth noting that similar values of the Curie-Weiss parameters were found in the previous studies on EuSnP \cite{Payne,Iha,Fujiwara}. As visualized in the inset to Fig. 1(b), $\chi(T)$ measured in a small magnetic field shows a very sharp peak at $T_N$ = 21(1) K that manifests the AFM phase transition. Another indication of the AFM character of the magnetic ground state in this compound comes from a fairly complex behavior of the magnetization isotherm taken at T = 2 K with magnetic field confined in the tetragonal plane. In concert with the literature data \cite{Iha,Fujiwara}, there occurs in $\sigma(H)$ a clear inflection near $\mu_0 H_m$ = 3.5(3) T, marking a metamagnetic-like transition, and another distinct anomaly in a field $\mu_0 H_{sat}$ = 6.1(1) T, above which a fully polarized spin state is gradually approached on further increase of the field strength (Fig. 1(c)).

The temperature dependence of the electrical resistivity of single-crystalline EuSnP, measured with electric current flowing within the tetragonal plane, is shown in Fig. 1(d). Worth noting is small magnitude of the resistivity in the entire temperature range covered, yet especially the residual resistivity of only 0.12 mWcm that gives rise to the residual resistivity ratio RRR = $\rho(300 K)$/$\rho(2 K)$ = 270, which is a value much larger than those reported before \cite{Payne, Iha}. Both parameters reflect very high crystalline quality of the EuSnP sample investigated. As can be inferred from Fig. 1(d), the onset of the AFM state manifests itself as a distinct kink in $\rho(T)$. Below $T_N$, the resistivity sharply decreases due to suppression of scattering conduction electrons on disordered magnetic moments. The most striking feature of $\rho(T)$ of EuSnP is its perfectly linear behavior in the entire PM state. The linear-in-T electrical transport was recognized as a characteristic property of strange metals (cuprates, heavy-fermion systems, etc.), in their quantum critical regimes, but also observed for some conventional metals (e.g., Li, Mg, Ag), where scattering from phonons is limited by the Plankian dissipation rate \cite{Planckian}. In order to elucidate the actual origin of the T-linear resistivity of EuSnP, further experimental and theoretical studies aimed at gaining detailed information about the charge carriers in this material are compulsory.\\
\indent Given the presence of superconductivity in sister compound SrSnP \cite{GuiSuperconducting2018}, our finding of linear-in-T resistivity behavior, and the presence of an AFM ground state, the electronic structure of EuSnP needs to be elucidated in order to determine the band structure's role in these phenomenon. To gain insight into the electronic structure of this material, we have presented the bulk Brillouin zone (Fig. 1(e)) and the electronic band dispersion along the corresponding high symmetry directions, as shown in Fig. 1(f). The yellow lines indicate the band dispersion without the inclusion of spin-orbit coupling (SOC), while the blue lines represent the dispersion with SOC. The calculated band structure shows that there are multiple bands which crosses the Fermi level (E$_F$) suggesting it to be a metal, in accordance with our transport measurements. In the absence of SOC, we can observe multiple band crossings along the $\Gamma$-X, M-$\Gamma$, $\Gamma$-Z-R, and A-Z direction. With the inclusion of SOC these band crossings become gapped. One seemingly important feature of the band structure occurs at the $\Gamma$-point, where a hole-like curvature of the valence band is seen along the $\Gamma$-X direction and the $\Gamma$-M direction, however, moving along $k_z$ in the $\Gamma$-Z direction shows electron-like curvature. Previous calculations on a related material, SrSnP, have attributed these features to the presence of a saddle-point at $\Gamma$ close to the Fermi level \cite{GuiSuperconducting2018}. To evaluate the validity of this description in EuSnP, we have performed density of states calculations, which shows no pronounced peak near the Fermi level (see the Supplemental Material \cite{Suppl}). Absence of a saddle point in EuSnP could explain the lack of low-temperature superconductivity in this material, contrasting with SrSnP in this regard. 

\begin{figure*}
	\includegraphics[width=18cm]{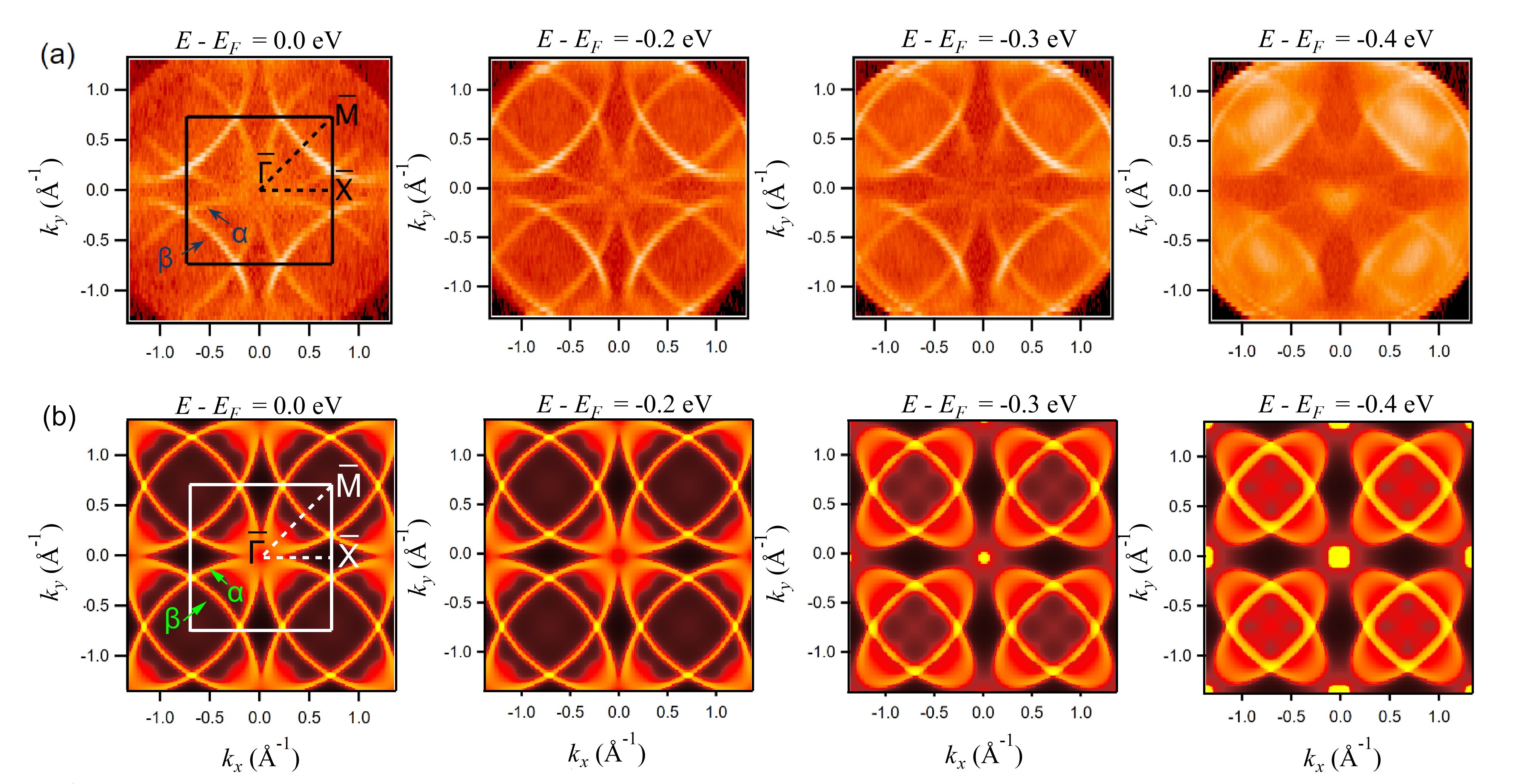}
	\caption{Fermi surface and constant energy contours. (a) ARPES measured Fermi surface (first panel) and constant energy contours measured using a photon energy of 100 eV at various binding energies as indicated on top of each plot. (b) Respective Fermi surface and constant energy contours obtained from DFT calculations. The experimental data were taken at the SSRL beamline 5-2 at a temperature of 18.5 K using LH polarization.}
	\label{fig2}
\end{figure*}

\indent We have measured the Fermi surface (FS) and constant energy contours (CECs) at various binding energies using a photon energy of 100 eV as shown in Fig. 2 in order to disclose the electronic structure of EuSnP. Fig. 2(a) shows the experimental FS maps and CECs, while Fig. 2(b) shows the DFT-calculated FS and CECs. Theoretical and experimental FSs are provided with BZs that have high-symmetry points marked on them. At the FS, several pockets are seen, pointing to the metallic band structure of the material. The experimental FS and the CECs of the material are quite well reproduced in the calculated plots shown in Fig. 2(b). The Fermi surface consists of two large concentric elliptical electron pockets surrounding the BZ corners. The four elliptical pockets oriented with their eccentric axis along the $\overline{\Gamma}$-$\overline{\text{M}}$ direction, which we label as $\alpha$, coincide at the $\overline{\Gamma}$ point. The interaction between these pockets produce a complex electronic structure which shows hole-like behavior along the $\overline{\Gamma}$-$\overline{\text{M}}$ direction, however which produces four electron-like pockets along the $\overline{\Gamma}$-$\overline{\text{X}}$ direction. Increasing the binding energy to 0.4 eV below the Fermi level shows the presence of a circular hole pocket surrounding the $\overline{\Gamma}$ point. Increasing the binding energy further adds another small circular pocket surrounding $\overline{\Gamma}$.

\begin{figure*}
	\includegraphics[width=18cm]{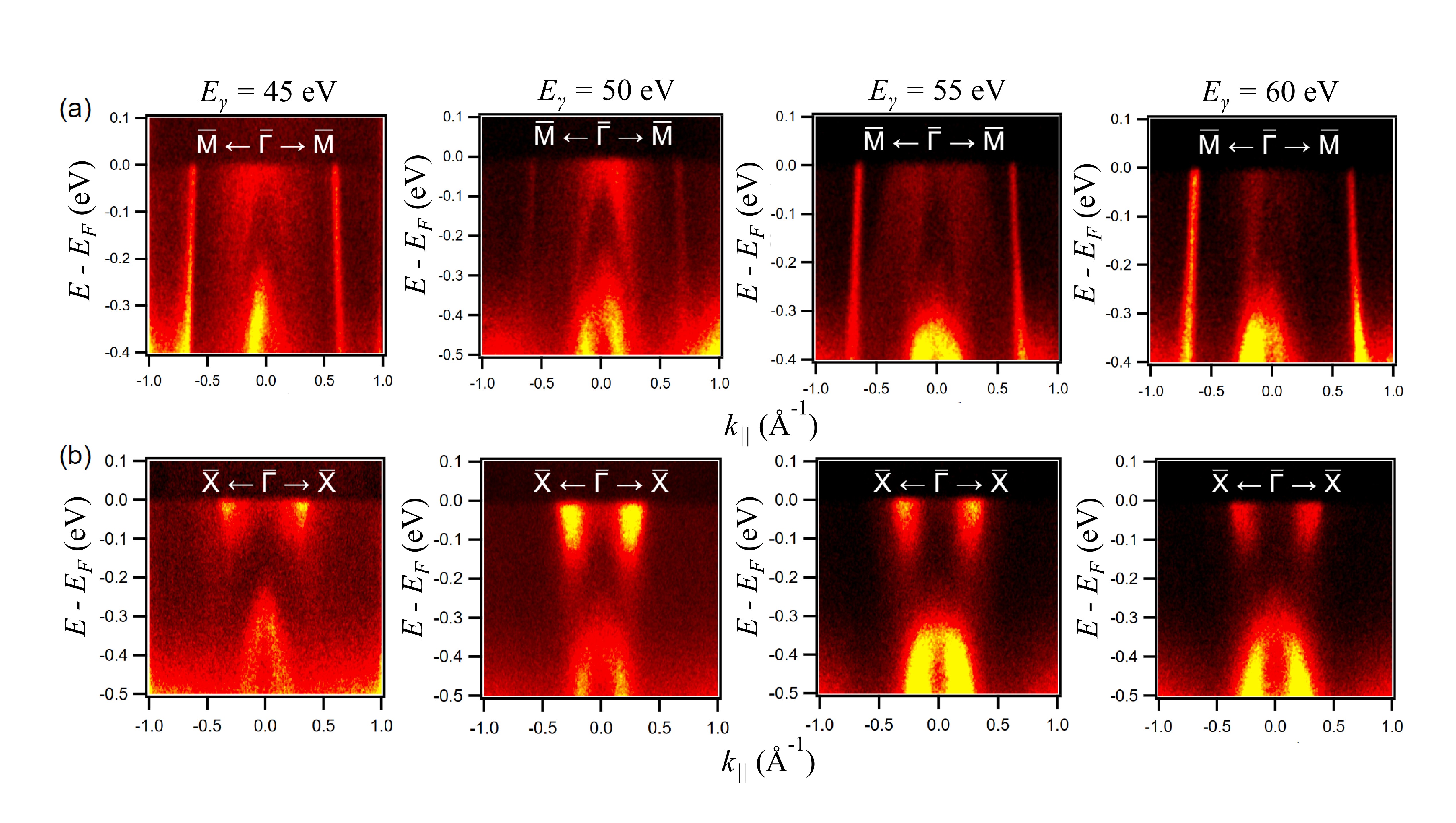}
	\caption{Band dispersion along the $\overline{\text{M}}$--$\overline{\Gamma}$--$\overline{\text{M}}$ and $\overline{\text{X}}$--$\overline{\Gamma}$--$\overline{\text{X}}$ directions. (a, b) Experimentally measured band dispersions along the $\overline{\text{M}}$--$\overline{\Gamma}$--$\overline{\text{M}}$ and $\overline{\text{X}}$--$\overline{\Gamma}$--$\overline{\text{X}}$ directions using photon energies of 45 eV, 50 eV, 55 eV, and 60 eV, respectively. The experimental data were taken at the SSRL beamline 5-2 at a temperature of 18.5 K using LH polarization.
	}
	\label{fig3}
\end{figure*}

\indent In order to understand the electronic structure of EuSnP at low temperatures, we have analyzed the band dispersion along various high-symmetry directions. Fig. 3(a) presents ARPES-obtained band dispersions along the $\overline{\text{M}}$-$\overline{\Gamma}$-$\overline{\text{M}}$ direction for photon energies ranging from 45 eV to 60 eV. A sharp steeply dispersing band is observed, crossing the Fermi level at about 0.5 $\text{\AA}^{-1}$, which is attributed to the co-vertex of the $\beta$ elliptical pocket. This band appears as a sharp, intense spectral feature, which does not appear to disperse over various photon energies (Fig. 3(a)). A second feature observed near the Fermi level is a hole-like band with relatively broad spectral features. Comparing with the Fermi surface, we attribute this hole pocket around $\overline{\Gamma}$ to the vertex of ellipse $\alpha$. This pocket disperses strongly as a function of photon energy. Another pocket below 0.2 eV at the $\overline{\Gamma}$ point strongly disperses with photon energy indicating it to be of bulk 3D origin. Next, we discuss the band dispersion measured along the $\overline{\text{X}}$-$\overline{\Gamma}$-$\overline{\text{X}}$ direction. The measured band dispersion with a photon energy of 45 eV suggests that there is no pocket at the $\overline{\Gamma}$ point. However an electron-like pocket exists along the $\overline{\Gamma}$-$\overline{\text{X}}$ direction. This pocket is highly dispersive as the photon energy is changed from 45 eV to 60 eV, suggesting its bulk origin. The pocket below 0.2 eV is strongly photon energy dependent indicating its bulk nature.

\begin{figure*}
	\includegraphics[width=18cm]{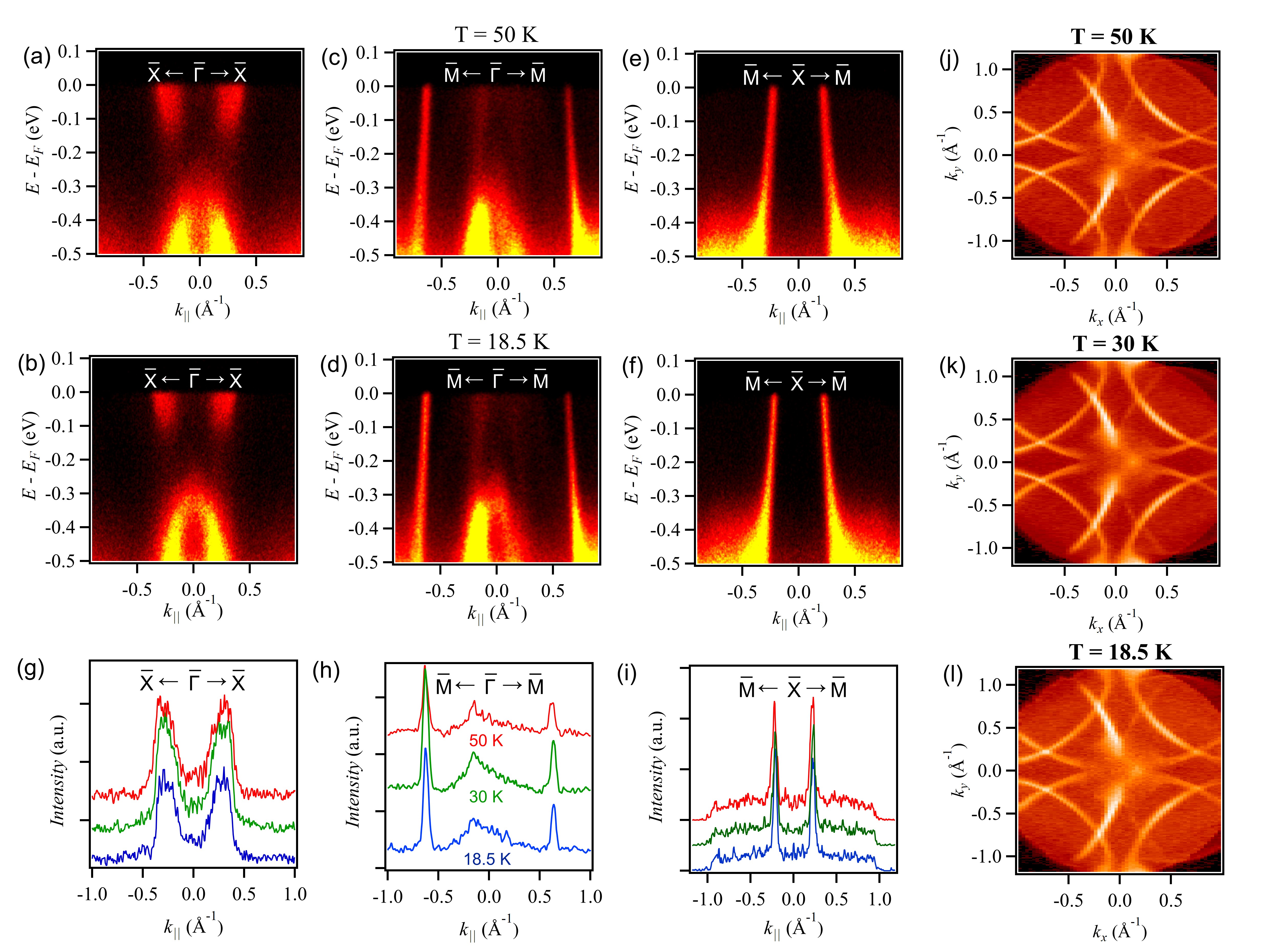}
	\caption{Temperature dependent band dispersion of EuSnP along various high-symmetry directions. (a, b) Experimentally measured band dispersions along the $\overline{\text{X}}$--$\overline{\Gamma}$--$\overline{\text{X}}$, (c, d) $\overline{\text{M}}$--$\overline{\Gamma}$--$\overline{\text{M}}$, and (e,f) $\overline{\text{M}}$--$\overline{\text{X}}$--$\overline{\text{M}}$ directions at 50 K and 18.5 K, respectively. (g-i) Momentum distribution curves (MDCs) at E$_F$ as a function of temperature. The high-symmetry directions are indicated in the captions. Comparison of the Fermi surface measured at (j) 50 K, (k) 30 K, and (l) 18.5 K, respectively. The experimental data were taken at the SSRL beamline 5-2 using LH polarization.
	}
	\label{fig4}
\end{figure*}

\indent To understand the effect of antiferromagnetism on the electronic band structure of EuSnP we have performed temperature dependent ARPES measurements along various high-symmetry directions as shown in Fig. 4. The measurements have been performed at 50 K, 30 K, and 18.5 K which is above and below the N\'eel temperature. Fig. 4(a-f) presents the ARPES measured band dispersion along the  $\overline{\text{X}}$-$\overline{\Gamma}$-$\overline{\text{X}}$, $\overline{\text{M}}$-$\overline{\Gamma}$-$\overline{\text{M}}$, and $\overline{\text{M}}$-$\overline{\text{X}}$-$\overline{\text{M}}$ directions at temperatures of 50 K and 18.5 K, respectively. The FS measured above and below the N\'eel temperature is presented in Fig. 4(j-l). As can be seen on comparing Fig. 4(a,b), Fig. 4(c,d), and Fig. 4(e,f), we do not observe any significant change in the electronic structure across the magnetic transition above and below the N\'eel temperature (see the Supplemental Material for further temperature dependent measurements \cite{Suppl}). In order to quantitavely analyze any change in the electronic structure as a function of temperature we have performed a close inspection of the momentum distribution curves (MDCs) at the Fermi level along the several high-symmetry directions, as shown in Fig. 4(g-i) which reveals 
the ARPES spectra does not demonstrate significant modification across the magnetic transition.

\indent Moving on to the interpretation of these results; we first note that the coupling between the magnetic \textit{f} electrons and conduction electrons define the modification to the electronic structure across the magnetic phase transition. Our present temperature dependent ARPES results suggest that \textit{f} electrons do not strongly interact with the valence bands and thus does not have any significant effect on the electronic structure of this material \cite{sobota, Yin}. As discussed in ref \cite{sobota}, one may justify several reasons for the lack of observed AFM reconstruction via ARPES. Firstly, the AFM ordering could conceivably be a strong bulk effect and may not extend to the surface. 
%However, in the context of PdCrO$_2$ studied in ref \cite{sobota}, a disagreement between low temperature band structure as measured by quantum oscillations and via ARPES was reported \cite{OkPrOscillations}. 
We note that EuSnP is revealed to show good agreement between the AFM-phase dHvA measurements \cite{Iha} and the obtained ARPES results, indicating that both the bulk and surface electronic structures see no reconstruction. The more likely scenario is where there is very weak coupling between the Eu moments and the Sn-P conduction electrons. This is understood to indicate that, while the AFM ordering does redefine the scale of the unit cell, the spectral weight associated with the AFM band folding is directly related to the coupling strength of the itinerant electrons to the Eu moments \cite{Voit2000Periodic}. Given the overall consistency between the previously reported bulk-sensitive dHvA measurements and the surface-sensitive ARPES measurements both above and below T$_N$, we associate the lack of observed band-folding, band-splitting, or band-shifts as an indication of the weak interaction strength between Eu 4$f$ electrons and the valence electronic states. 

\indent In conclusion, we have successfully grew high-quality single crystals of EuSnP, which possess a tetragonal space group of P4\textit{/nmm}. Our investigations revealed an antiferromagnetic (AFM) transition in this material, with a N\'eel temperature of 21 K, as evidenced by magnetic measurements. Resistivity measurements indicated EuSnP to be a good electrical conductor, showing a striking linear temperature dependence of the resistivity in the PM phase. In order to explore the electronic structure of this compound, we employed ARPES (angle-resolved photoemission spectroscopy). Our ARPES measurements unveiled significant elliptical pockets centered around the $\overline{\text{M}}$ point, corresponding to the conduction electrons. Additionally, we observed small bulk electron pockets along the $\overline{\Gamma}$--$\overline{\text{X}}$ direction, as well as small hole-pockets along the $\overline{\Gamma}$--$\overline{\text{M}}$ direction. Importantly, our experimental results were well reproduced by DFT-based first principles calculations.  Surprisingly, upon cooling the sample below the N\'eel temperature, we observed no electronic reconstruction. This suggests that the Sn and P electrons weakly couple to the magnetic order of the Eu spins. These techniques will shed light on the interplay between the Sn and P conduction electrons and the Eu spins. Our work not only provides a comprehensive framework for understanding the electronic structure of this new material both above and below the N\'eel temperature but also serves as a valuable resource for future researchers interested in exploring the effects of antiferromagnetism in EuSnP.\\

\indent M.N. is supported by the Air Force Office of Scientific Research MURI (Grant No. FA9550-20-1-0322), and the National Science Foundation (NSF) CAREER award DMR-1847962. D.K. and T.R. were supported by the National Science Centre (NCN, Poland) under research grant 2021/41/B/ST3/01141.  A.P. acknowledges the support by National Science Centre (NCN, Poland) under Projects No. 2021/43/B/ST3/02166. The use of Stanford Synchrotron Radiation Lightsource (SSRL) in SLAC National Accelerator Laboratory is supported by the U.S. Department of Energy, Office of Science, Office of Basic Energy Sciences under Contract No. DE-AC02-76SF00515. We thank Makoto Hashimoto and Donghui Lu for the beamline assistance at SSRL endstation 5-2. This research also used resources of the ALS at the Lawrence Berkeley National Laboratory, a US Department of Energy Office of Science User Facility, under Contract No. DE-AC02-05CH11231. We thank Jonathan Denlinger for beamline assistance at the ALS Beamline 4.0.3.\\
 
%\begin{thebibliography}
%\noindent \textbf{REFERENCES}
%\def\bibsection{\section*{\refname}}

%{\noindent \textbf{AUTHOR CONTRIBUTIONS}\\
%M.N. and D.K. conceived the project; T.R. synthesized the crystals and tested their quality by XRD, EDX, magnetic, and transport measurements; A.P.S performed the ARPES measurements with the help of M.S., S.R., M.I.M., and I.B.E.; M.S. and A.P.S analyzed the data and wrote the manuscript with the input from all the authors; A.P. performed the theoretical calculations; M.N. was responsible for the overall research direction, planning, and integration among different research units; All authors discussed the results, interpretation and conclusion.\\

%\noindent \textbf{Data availability}\\
%The data supporting the findings of this study are available within the paper, and other findings of this study are available from the corresponding author upon reasonable request.\\

%\noindent \textbf{Competing interests}\\
%The authors declare no competing interests.\\

%\noindent \textbf{ADDITIONAL INFORMATION}\\
%\textbf{Correspondence} and requests for materials should be addressed to Madhab Neupane.
\end{document}